\documentclass{ws-p10x7}
\input babarsym
\newcommand{\aerr}[4]   {\mbox{${{#1}^{+ #2}_{- #3}\pm #4}$}}
\newcommand{\berr}[4]   {\mbox{${{#1}\pm #2^{+ #3}_{- #4}}$}}
\newcommand{\cerr}[3]   {\mbox{${{#1}^{+ #2}_{- #3}}$}}
\newcommand{\aerrsy}[5] {\mbox{${{#1}^{+ #2 + #4}_{- #3 - #5}}$}}

\newcommand{\berrsyt}[6] {\mbox{${{#1}\pm #2^{+ #3 + #5}_{- #4 - #6}}$}}
\newcommand{\err}[3]   {\mbox{${{#1}\pm{#2}\pm{#3}}$}}

\begin{document}

\title{CKM Parameters and Rare B Decays}

\author{F. Forti}

\address{INFN-Pisa, L.go Pontecorvo, 3, 56127 Pisa, Italy\\
E-mail: Francesco.Forti@pi.infn.it}
\report{SLAC-PUB-11543}

\twocolumn[\maketitle\abstract{
Measurements of the angles and sides of the unitarity triangle and of
the rates of rare $B$ meson decays are crucial for the precise
determination of Standard Model parameters and are sensitive to the
presence of new physics particles in the loop diagrams. In this paper
 the recent measurements performed in this area by
\babar\ and Belle will be presented.
The direct measurement of the angle $\alpha$ is for the first time as
precise as the indirect determination.  The precision of the
$|V_{ub}|$ determination has improved significantly with respect to
previous measurement. New limits on $B\to\tau\nu$ decays are
presented, as well as updated measurements on $b\to s$ radiative
transitions and a new observation of $b\to d\gamma$ transition made by
Belle.

}]

\section{Introduction}
\begin{figure}
\epsfxsize180pt
\figurebox{}{}{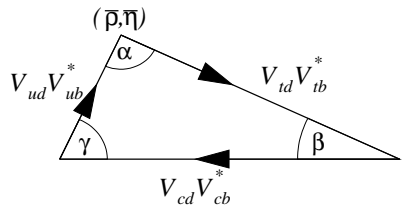}
\caption{Unitarity triangle.}
\label{fig:ut}
\end{figure}

In the Standard Model (SM), the interaction between up-type and
down-type quarks is described by a unitarity matrix called the
Cabibbo-Kobayashi-Maskawa matrix (in brief
CKM).\cite{Cabibbo:1963yz,Kobayashi:1973fv} This matrix can be
parametrized with 3 real angles and one complex phase, which gives
rise to CP violation. A widely used parametrization of the
matrix\cite{Wolfenstein:1983yz,Buchalla:1995vs} uses the four parameters
$A,\lambda,\rhobar,\etabar$, with $\etabar$ controling the CP violation in this
framework. The unitarity of the CKM matrix imposes 9 complex relations
amongst the matrix elements, one of which is given by
\[
V_{ub}^*V_{ud}^{\phantom{*}}+V_{cb}^*V_{cd}^{\phantom{*}}+V_{tb}^*V_{td}^{\phantom{*}} =0,
\]
where $V_{qq\prime}$ is the matrix element relating the quark $q$ and $q\prime$.
This relation can be represented as a triangle 
(called the unitarity trangle) in the complex $\rhobar,\etabar$ 
plane, as shown in Fig.~\ref{fig:ut}.  $B$-meson decays are sensitive
probes to measure both the angles and sides of the unitarity triangle
and can unveil physics beyond the SM. In fact, most $B$
decay amplitudes receive contributions from diagrams containing loops,
where the presence of new particles can be detected through effects on
the branching ratios, asymmetries, or spectra. Another possible route
to detecting new physics is the high precision measurement of the
unitarity triangle parameters to uncover any inconsistency 
among them or between different determinations of the same parameter.

After having clearly established CP-violation in the $B$ sector, the
\babar\ and Belle experiments are now pursuing an extended program of
precision measurements of the unitarity triangle parameters and of
rare $B$ decays, taking advantage of the very large data sample
collected at the B-Factories. The recent results of this measurement
program are reported at this conference in two papers.  The
measurement of $\sin 2\beta$ and the direct CP violation measurements
are presented by Kazuo Abe.  In this paper, after introducing the
\babar\ and Belle experiments in Sec.~\ref{sec:exp}, I will cover the
$\alpha$ measurements in Sec.~\ref{sec:alpha} and the $|V_{ub}|$ and
$|V_{cb}|$ measurements in Sec.~\ref{sec:vub}. The rest of the paper
will be devoted to rare decays: $\B\to\tau\nu$ in
Sec.~\ref{sec:btaunu}, $b\to s\gamma$ in Sec.~\ref{sec:bsgamma}, and
$b\to d\gamma$ in Sec.~\ref{sec:bdgamma}. I will finally give some
concluding remarks in Sec.~\ref{sec:summary}.

\section{The B-Factory experiments and datasets}
\label{sec:exp}
The data used in the analyses presented in this paper have been
collected with the \babar\ detector at the PEP-II machine at SLAC and
with the Belle detector at the KEKB machine in the KEK laboratory
between 1999 and 2005. Both detectors, whose detailed description can
be found elsewhere,\cite{Aubert:2001tu,belle:2002cg} have been
designed and optimized to study time-dependent CP-violation in $B$
decays at the $\Upsilon(4S)$ resonance. Their major components are: a
vertexing and tracking system based on silicon and gas detectors; a
particle identification system; an electromagnetic calorimeter based
on CsI(Tl) crystals operating within a 1.5~T magnetic field; an iron
flux return located outside of the coil, instrumented to detect $K^0_L$
and identify muons. The $\Upsilon(4S)$ resonance decays most of the
time in a pair of $B$-mesons, either \BpBm or \BzBzb, which acquire a
boost thanks to the asymmetry of the beam energies: $9~\gev~e^-$ on
$3.1~\gev~e^+$ for PEP-II and $8~\gev~e^-$ on $3.5~\gev~e^+$ for
KEKB. Because of this boost, the decay vertices of the two mesons
are  separated, thus allowing their individual determination and
the measurement of time-dependent CP asymmetries. In these analyses,
the signal $B$ is reconstructed in a CP-eigenstate (such as
$B\to\pi\pi$) while the other $B$ (the tagging $B$) is reconstructed
in a decay mode that allows the determination of its flavor at the
time of decay, such as exclusive hadronic or semileptonic modes, or
inclusive modes with a lepton or a kaon, whose sign carries the
information of the $B$ flavor.

The data samples used in the measurements presented in this paper vary
for the two experiments. Most measurements are based on $232\times
10^{6}\BB$ pairs for \babar\ and $275\times 10^{6}\BB$ for
Belle, but there are several results obtained with smaller statistics,
while Belle performed the $b\to d\gamma$ analysis with $385\times 10^{6}\BB$.

\section{Determination of the angle $\alpha(\Phi_2)$}
\label{sec:alpha}

\begin{figure}
\epsfxsize200pt
\figurebox{}{}{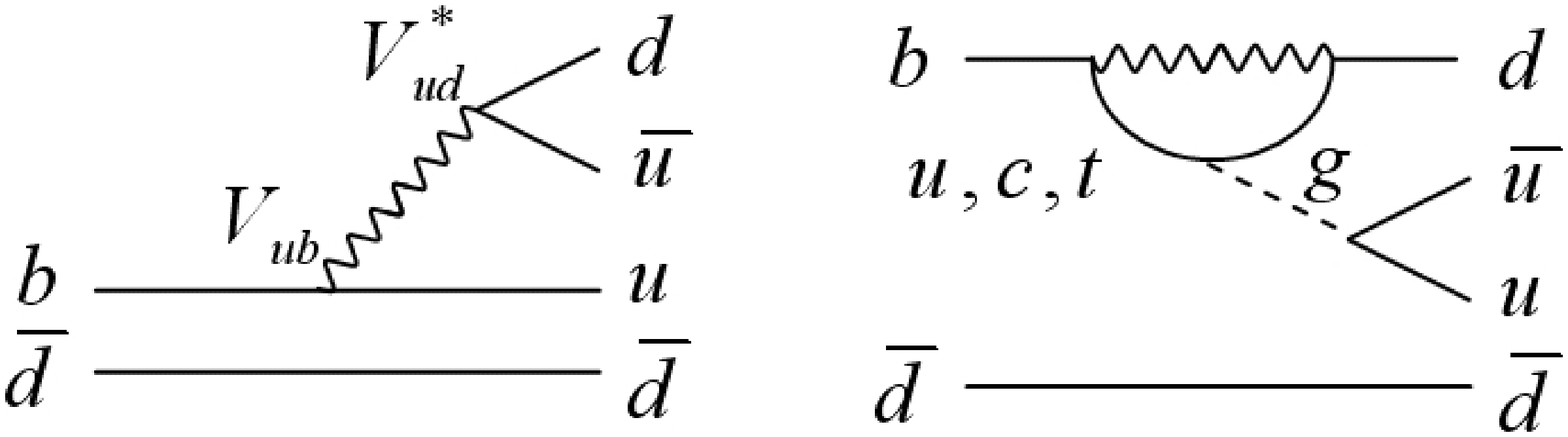}
\caption{The tree (left) and penguin (right) diagrams contributing 
     to charmless $B$ decays \Bztopipi, \Bztorhopi and \Bztorhorho.}
\label{fig:chmlsdiagrams}
\end{figure}

The angle $\alpha$ is the relative phase of the $V_{ub}$ and $V_{td}$ CKM
matrix elements and can be measured in the charmless $B$ decays
$B\to\pi\pi$, $B\to\rho\pi$ and $B\to\rho\rho$ which arise from tree-level 
$b\to u(\overline u d)$ transitions (Fig.~\ref{fig:chmlsdiagrams},left). 
A complication to this approach is the presence of loop level
penguin diagrams leading to the same final states
(Fig.~\ref{fig:chmlsdiagrams}, right), which introduce different CKM
matrix elements.
While in the absence of penguin contribution, the measurement of time
dependent CP asymmetries in neutral $B$ charmless decays would
directly yield the angle $\alpha$, the interference between tree 
and penguin diagrams
obscures the simple relationship between CP observables and the angle
$\alpha$ and requires the development of specific techniques to
disentangle the penguin contribution.

Time-dependent CP asymmetries arise
from the intereference of two possible paths reaching the same final
state: $B\to f$ and $B\to\overline B\to f$, and can be expressed in
terms of the complex parameter 
$\lambda_f = \eta_f\frac{p}{q}\frac{\overline A}{A}$, where 
$A=|\langle f|T|\Bz \rangle|$, $\overline A=|\langle f|T|\Bzb\rangle|$, 
$\eta_{f}$ is the CP eigenvalue of the final state and $q,p$ are
the parameters describing how \Bz and \Bzb mix to form the mass
eigenstates. The time dependent CP asymmetry follows
\[
A_{CP}(\Delta t) = S_f \sin(\Delta m \Delta t) + C_f \cos(\Delta m \Delta t),
\]
where $S_f = 2\frac{\Im (\lambda_f)}{1+|\lambda_f|^2}$ measures the 
CP violation arising from the
interference of the decays with and without mixing, and
$C_f = \frac{1-|\lambda_f|^2}{1+|\lambda_f|^2} $ 
measures the direct CP violation in the decay. For the tree diagram in 
(Fig.~\ref{fig:chmlsdiagrams},left) 
\[
\lambda_f=\eta_f\frac{V_{tb}^*V_{td}V_{ub}V_{ud}^*}{V_{tb}V_{td}^*V_{ub}^*V_{ud}}=\eta_f e^{2i\alpha},
\]
with $C_f=0$ and $S_f=\sin(2\alpha)$. In the presence of the penguin
diagram, the expression becomes:
\[
\lambda_f=\eta_f e^{2i\alpha}\frac{T+Pe^{+i\gamma}e^{i\delta}}{T+Pe^{-i\gamma}e^{i\delta}}
\]
where $T$ and $P$ are the tree and penguin amplitudes, and $\delta$ is
the strong phase. The effect of penguin diagram interference is the
possibility of direct CP violation ($C_f\propto\sin\delta$) and a
shift $\Delta\alpha$ in the measurement of the angle $\alpha$:
$S_f=\sqrt{1-C_f^2}\sin(2\alpha_{\rm eff})$ with
$\Delta\alpha=\alpha_{\rm eff}-\alpha$.

\begin{figure}
\epsfxsize180pt
\figurebox{}{}{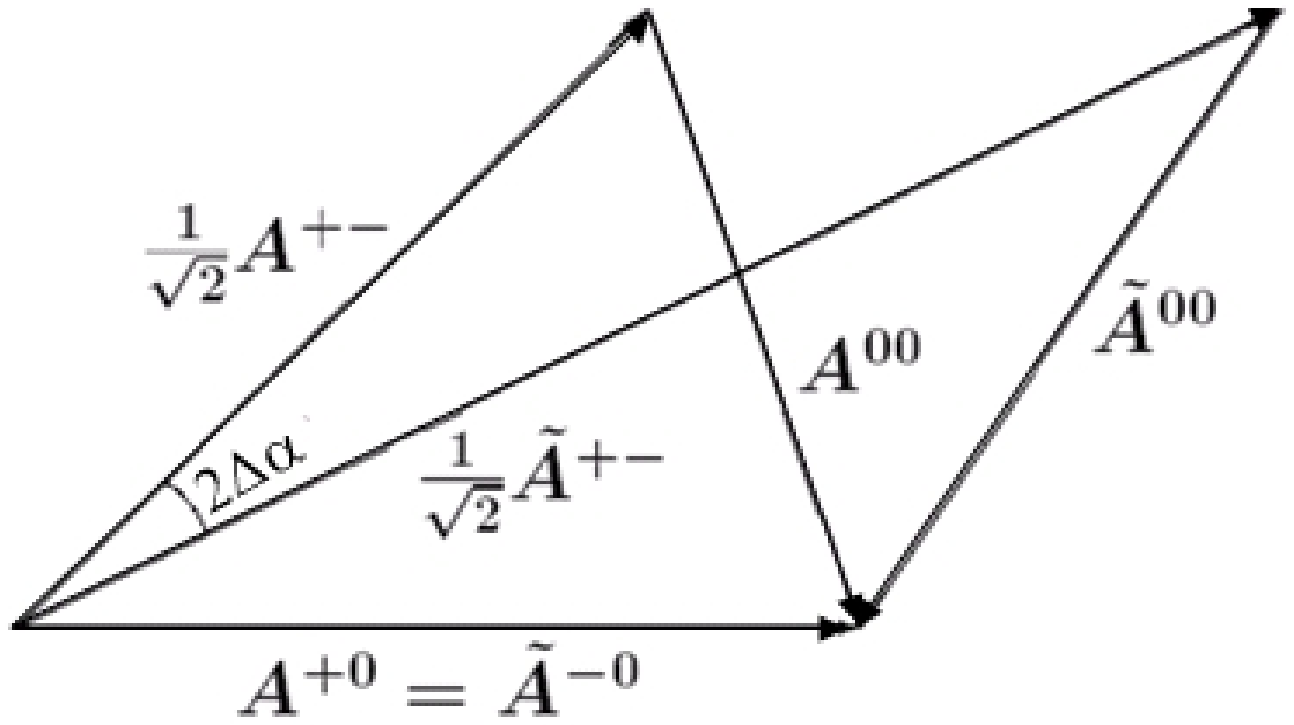}
\caption{Isospin triangles for the charmless $B$ decays \Bztopipi, \Bztorhorho.}
\label{fig:isospin}
\end{figure}

Isospin relations amongst rates of the various $\B\to\pi\pi$
and $\B\to\rho\rho$ decays can be used\cite{Gronau:1990ka} to extract
the shift $\Delta\alpha$. The isospin analysis involves the separate
measurement of \Bz and \Bzb decay rates into $h^+ h^-$ ($h$ indicates
either $\pi$ or $\rho$) and $h^0 h^0$, as well as the measurement of
the rate of the charged $B$ decay $\B^{+(-)}\to h^{+(-)} h^0$. Constructing a \Bz
and a \Bzb triangle from the 6 amplitudes (Fig.~\ref{fig:isospin}) one
can extract $\Delta\alpha$ from the mismatch of the two
triangles.

\begin{figure}
\epsfxsize200pt
\figurebox{}{}{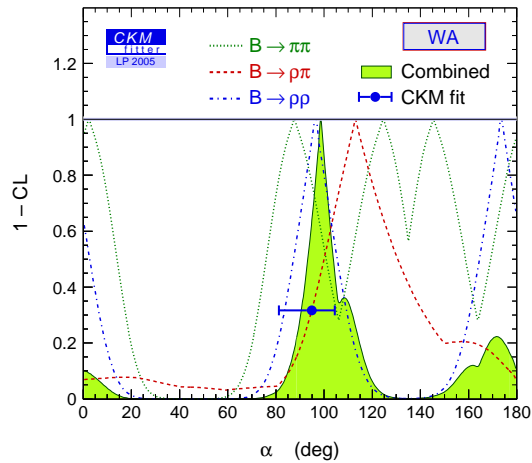}
\caption{Alpha determination from the charmless $B$ decays \Bztopipi, \Bztorhorho and \Bztorhopi.
The dotted lines represent the results of the three individual
analyses. The green area is the result of the combined fit. The CKM
triangle fit independent determination of alpha, which is not included
in the fit, is shown by the blue point.}
\label{fig:alpha}
\end{figure}

It has also been shown\cite{Grossman:1997jr} that, in alternative to
full isospin analysis, one can use the branching fractions for 
$\B\to h^0 h^0$ and $\B\to h^+ h^0$ averaged over meson and anti-meson to 
impose an upper bound on $\Delta\alpha$:
\[
\sin^2\Delta\alpha < \frac{ \overline\BR(\Bz\to h^0 h^0)}{\overline\BR(\Bpm\to h^\pm h^0)}
\]
Other relations have also been
developed,\cite{Charles:1998qx,Gronau:2001ff} but with the current
level of accuracy of the measurements none improves significantly over
the above limit.
The constraints on $\alpha$ derived from a full isospin analysis 
in the $\pi\pi$ channel\cite{Aubert:2005av,Abe:2005dz}
are very weak, as shown in Fig.~\ref{fig:alpha} explained later in
the text, mainly
due to the fact that the branching ratio 
$\BR(\Bz\to\pi^0 \pi^0)=(1.45 \pm 0.29)\times 10^{-6}$ 
(averaged by HFAG\cite{HFAG} on the basis of the
\babar\cite{Aubert:2004aq} and Belle\cite{Abe:2004mp} measurements) 
is too large to be effective in setting the above limit, but
is also too small for the full isospin analysis.

\begin{figure}
\epsfxsize200pt
\figurebox{}{}{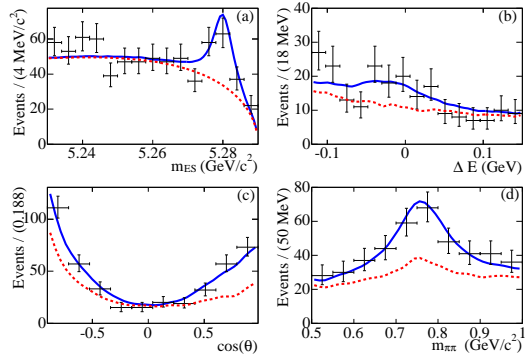}
\caption{The distributions for the highest purity tagged events 
      in the \babar\ $B\to\rho\rho$ analysis for the variables $m_{ES}$ (a),
      $\Delta E$ (b), cosine of the $\rho$ helicity angle (c), and
      $m_{\pi^\pm\pi^0}$ (d). The dotted lines are the sum of
      backgrounds and the solid lines are the full PDF.}
\label{fig:rhorhormES}
\end{figure}

The $\rho\rho$ channel has three polarization amplitudes, which
introduce dilution in the measurement because they have different CP
eigenvalues, and has been considered in the past as less promising
than $\pi\pi$. Both \babar\ and Belle have recently performed full
analyses of this decay.\cite{Aubert:2005nj,Abe:2005ft} The charged
$\rho$ is reconstructed through the decay $\rho^\pm\to\pi^\pm\pi^0$,
and the events are selected through a kinematical signal
identification based on the beam-energy substituted mass (also known
as beam constrained mass) $m_{\rm bc} \equiv m_{\rm ES}=\sqrt{E_{\rm
beam}^{*2}-p_{B}^{*2}}$ and the energy difference between the
reconstructed $B$ and the beam $\Delta E = E_{B}^{*} - E_{\rm
beam}^{*}$. All quantities  are computed in the CM
frame. The distribution of these variables for the signal and the
background is shown in Fig.~\ref{fig:rhorhormES}

\begin{table*}[tb]
\caption{Summary of measurements for the $\B\to\rho\rho$ decays}
\label{tab:rhorho}
\begin{center}
\begin{tabular}{|c|c|c|} 
 
\hline 
Quantity & \babar & Belle \\\hline
\raisebox{0pt}[16pt][6pt]{$f_L$} &
$\berr{0.978}{0.014}{0.021}{0.029}$ &
$\aerrsy{0.951}{0.033}{0.039}{0.029}{0.031}$
\\\hline

\raisebox{0pt}[16pt][6pt]{$S_{\rho\rho,L}$} &
$\berr{-0.33}{0.24}{0.08}{0.14}$ &
$\err{0.09}{0.42}{0.08}$
\\\hline

\raisebox{0pt}[16pt][6pt]{$C_{\rho\rho,L}$} &
$\err{-0.03}{0.18}{0.09}$ &
$\berr{0.00}{0.30}{0.09}{0.10}$ 
\\\hline

\raisebox{0pt}[16pt][6pt]{$\BR(\Bz\to\rho^+\rho^-)$ $[10^{-6}]$ } &
$\err{30}{4}{5}$                            & 
$\berr{24.4}{2.2}{3.8}{4.1}$                 
\\\hline

\raisebox{0pt}[16pt][6pt]{$\BR(\Bpm\to\rho^+\rho^0)$ $[10^{-6}]$ }&
$\aerr{22.5}{5.7}{5.4}{5.8}$                      & 
$\berr{31.7}{7.1}{3.8}{6.7}$                        
\\\hline

\raisebox{0pt}[16pt][6pt]{$\BR(\Bz\to\rho^0\rho^0)$  $[10^{-6}]$} & $<1.1$ & - \\\hline

\end{tabular}
\end{center}
\end{table*}

It is found that the fraction of longitudinal polarization ($f_L$) in
the $\rho\rho$ final state is almost 100\%, and that therefore there
is no dilution effect in the measurement of $\alpha$. In addition, the
$\rho^+\rho^-$ and $\rho^+\rho^0$ branching fractions are a factor of
5 larger than the corresponding ones in the $\pi\pi$ decays, but at
the same time the $\rho^0\rho^0$ is not yet observed, with a
relatively small limit on $\Delta\alpha$. The results are summarized
in Table~\ref{tab:rhorho}

Using the \babar\ limit on $\BR(\Bz\to\rho^0\rho^0)$ and the average
between the two experiments for the other quantities one arrives at a
relatively stringent limit on $\Delta\alpha$ ($\Delta\alpha<11^\circ$)
and at the  determination $\alpha[\rho\rho]=(96\pm 13)^\circ$.

The isospin analysis has an intrinsic two-fold ambiguity that can be
removed with a full time-dependent Dalitz plot analysis of the
$B\to\rho\pi$ decay.\cite{Snyder:1993mx} Results on this analysis have been presented the
ICHEP04 conference.\cite{Wang:2004va,Aubert:2004iu}

The results of the three analysis are summarized in
Fig.~\ref{fig:alpha}, where a combined fit\cite{Charles:2004jd} is
also shown. The result of this combined fit is
$\alpha=(\cerr{99}{12}{9})^\circ$. The result from the indirect
measurement of $\alpha$ obtained by fitting all the other CKM triangle
measurements, $\alpha{\rm [CKM]}=(\cerr{96}{11}{12})^\circ$ is shown
for comparison on the same plot. This is the first time that the
direct measurement of $\alpha$ has a better precision than the its
indirect determination from the CKM triangle fit.


\section{Measurement of $|V_{ub}|$ and $|V_{cb}|$}
\label{sec:vub}

The magnitude of the CKM matrix elements $V_{ub}$ and $V_{cb}$ can be
extracted from the semileptonic decay rate of $B$ mesons. At the
parton level the decay rates for $b\to u\ell\nu$ and $b\to c\ell\nu$
can be calculated accurately; they are proportional to $|V_{ub}|^2$
and $|V_{cb}|^2$, respectively, and depend on the quark masses, $m_b$,
$m_u$, and $m_c$.  To relate measurements of the semileptonic decay
rate to $|V_{ub}|$ and $|V_{cb}|$, the parton-level calculations have
to be corrected for effects of strong interactions, thus introducing
significant theoretical uncertainties for both  exclusive and
inclusive analyses.

For of exclusive decays, the effect is parametrized by form
factors(FF), such as in the simple case of the $B\to\pi\ell\nu$ decay,
neglecting the $\pi$ mass:
\[
  \frac{d\Gamma(\Bz\to\pim\ellp\nu)}{dq^2} =
  \frac{G_F^2\Vub^2}{24\pi^3}|f_+(q^2)|^2p_{\pi}^3,
\]
where $G_F$ is the Fermi constant, $q^2$ is the invariant-mass squared
of the lepton-neutrino system and $p_\pi$ is the pion momentum in the
$B$ frame. The FF $f_+(q^2)$ can be calculated with a variety of
approaches based on quark model,\cite{ISGW2} Light Cone Sum
Rules,\cite{Ball05} and lattice QCD.\cite{HPQCD04,FNAL04} In inclusive
decays, the main difficulty is to relate the partial rate obtained by
the experimental event selection process to the matrix elements.  This
is a particularly serious issue for $|V_{ub}|$, where only a small
fraction of the total rate can be determined experimentally because of
the severe background rejection cuts.
Heavy-Quark Expansions (HQEs)\cite{HQE}
have become a useful tool for calculating perturbative and
non-perturbative QCD corrections and for estimating their
uncertainties. These expansions contain parameters 
such as the $b$ quark mass and the average Fermi momentum
of the $b$ quark inside the $B$ meson. These parameters must be
determined experimentally, for instance from the photon energy
spectrum in $B\to X_s\gamma$ decays and the spectrum of the hadronic
mass in $B\to X_c\ell\nu$ decays.

For the determination of $|V_{cb}|$, a global analysis of inclusive $B$
decays has been performed,\cite{Bauer:2004ve} leading to a very precise measurement:
$$
|V_{cb}|_{\rm incl.} = (\err{41.4}{0.6_{\rm exp}}{0.1_{\rm th}})\times 10^{-3}.
$$
The measurement obtained from the world average of 
$\BR(B\to D^*\ell\nu)$,\cite{Group(HFAG):2005rb}
$$
|V_{cb}|_{D^*\ell\nu} = (\err{41.3}{1.0_{\rm exp}}{1.8_{\rm th}})\times 10^{-3},
$$
is fully compatible, although less precise.

These accurate measurements demonstrate the rapid experimental and
theoretical advancements in these area.

\subsection{$b\to u\ell\nu$ inclusive decays.}
Several methods have been used to isolate inclusive $b\to u\ell\nu$
decays from the much more frequent $b\to c\ell\nu$ decays.

In the lepton endpoint method\cite{Limosani:2005pi,Aubert:2004bv} one
uses the fact that, due to the mass difference between $c$ and $u$
quarks, the lepton spectrum in the $b\to u$ transition extends to
slightly higher energies than in the $b\to c$ decays. The lepton
momentum window is typically $1.9<p_{\rm lept}<2.6\gevc$, and a
selection is applied on the basis of event shape variables and missing
momentum. The background remains in any case significant with
typically $S/B\approx 1/14$.

\begin{figure}
\epsfxsize200pt
\figurebox{}{}{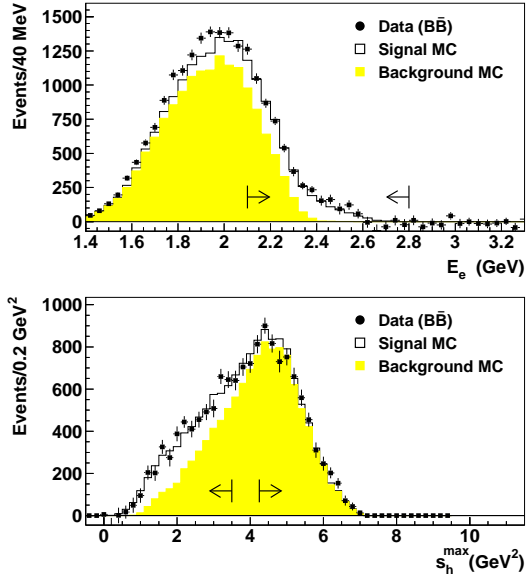}
\caption{The electron energy, $E_e$, and $s_h^{\rm max}$ spectra 
in the $\Upsilon(4S)$ frame for continuum-subtracted data and
simulated \BB\ events satisfying all the selection criteria except for
the variable shown. The arrows denote the signal and sideband
regions.}
\label{fig:vubq2}
\end{figure}

One can refine the selection by using a $q^2$-dependent electron
energy cut (the $E_e-q^2$ method)\cite{Aubert:2005im} where the
neutrino momentum is estimated from the event missing momentum and
$q^2$ is calculated from $q^2=(p_e+p_\nu)^2$. For each $E_e$ and $q^2$
one can calculate the maximum kinematically allowed hadronic mass
square $s_h^{\rm max}$ and veto $b\to c\ell\nu$ decays by requiring
$s_h^{\rm max}<3.5\gev^2\approx m^2_D$. This technique significantly
improves the S/B ratio to about 1/2. Figure~\ref{fig:vubq2} shows the
electron energy and $s_h^{\rm max}$ spectra, along with signal and
sideband regions.

Reconstructing the other $B$ in the event in an exclusive channel
allows the direct reconstruction of the hadronic system (called $X$)
produced in $b\to u\ell\nu$ decays by assigning all the remaining
particles to it. \babar\ uses the mass of the hadronic 
system to perform a 2-dimensional fit for the partial branching
fraction in the area $\{M_X<1.7\gevcc,\,
q^2>8\gev^2\}$,\cite{Aubert:2005hb} while Belle also introduces the
variable $P_+\equiv E_X - |{\bf p}_X|$, where $E_X$ and ${\bf p}_X$
are the energy and 3-momentum of the hadronic system, analyzing data
in three kinematical regions $M_X<1.7\gevcc$, $\{M_X<1.7\gevcc,
q^2>8\gev^2\}$, and $P_+<0.66\gevc$.\cite{Bizjak:2005hn}

\begin{figure}
\epsfxsize200pt
\figurebox{}{}{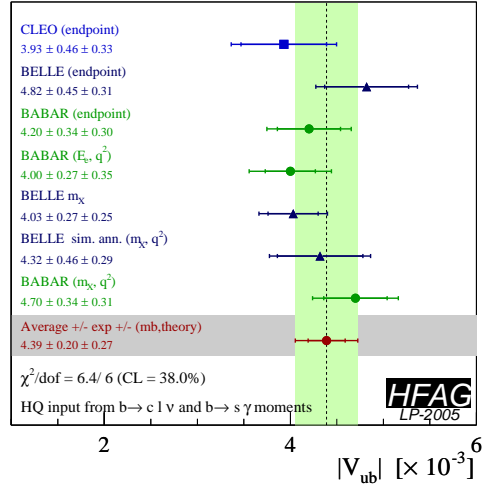}
\caption{Summary and average of inclusive $|V_{ub}|$ 
determinations using HQE parameters extracted from 
$B\to X_s\gamma$ and $B\to X_c\ell\nu$ moments.}
\label{fig:vub_allMoments}
\end{figure}

The extraction of $|V_{ub}|$ from these partial branching fractions
involves the determination of HQE parameters, which can be
done following a variety of schemes and using different physical
processes.\cite{Lange:2005yw} This extraction is the object of a very
active discussion with the goal of improving the precision of the
measurement.  A summary of $|V_{ub}|$ inclusive determinations based
on HQE parameters derived from the moments of the photon
energy spectrum in $B\to X_s\gamma$ decays and from the hadronic-mass
and lepton-energy moments in $B\to X_c\ell\nu$ decays is shown\cite{HFAG} in
Fig.~\ref{fig:vub_allMoments}:
$$
|V_{ub}|_{\rm incl.} = (\err{4.39}{0.20_{\rm exp}}{0.27_{\rm th}})\times 10^{-3}.
$$

An alternative determination, using HQE
parameters\cite{Bizjak:2005nk} obtained fitting the Belle $B\to
X_s\gamma$ photon energy spectrum, yields:\cite{Limosani:2005pi}
$$
|V_{ub}|_{\rm incl.} = (\err{5.08}{0.47_{\rm exp}}{0.48_{\rm th}})\times 10^{-3}.  
$$

\subsection{$B\to \pi\ell\nu,\rho\ell\nu$ decays.}
Various methods have been devised to isolate exclusive $B\to
\pi\ell\nu,\rho\ell\nu$ decays from the large backgrounds from $b\to
c\ell\nu$ and continuum events. Estimating the neutrino momentum from
the missing momentum in the event allows the usage of the mass of the
$B$ candidate $m_{ES}$ as a discriminating variable. In addition, one
can analyze the data in bins of $q^2$ (three bins for
CLEO\cite{Athar:2003yg} and five bins for \babar\cite{Aubert:2005cd})
and measure the $q^2$ dependance of the form factor, thus
discriminating among theoretical models.

Tagging the other $B$ in the event is another powerful method to
reduce backgrounds. As in the case of inclusive decays one can
reconstruct the other $B$ in an exclusive hadronic
channel\cite{Aubert:2004bq} (BReco tag) which allows the
reconstruction of the hadronic system on the signal side.
Alternatively, one can tag the other $B$ through semileptonic decays,
and use the kinematics of 2 back-to-back semileptonic decays to reduce
the background.\cite{Aubert:2005tm,Aubert:2005tn,Abe:2004zm}

\begin{figure}
\epsfxsize200pt
\figurebox{}{}{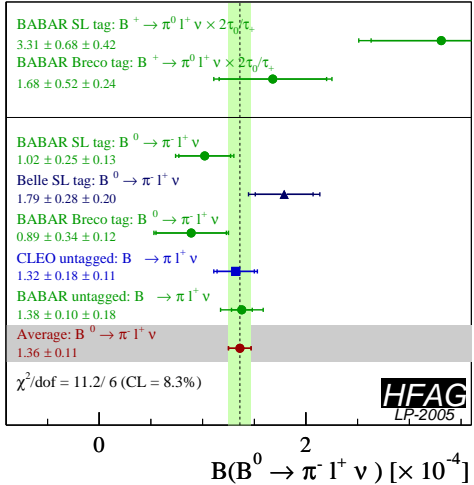}
\caption{Summary and average of exclusive $B\to\pi\ell\nu$ branching fractions.}
\label{fig:pilnu}
\end{figure}

A summary of exclusive $B\to\pi\ell\nu$ branching fractions
measurements is shown in Fig.~\ref{fig:pilnu}.  The extraction of
$|V_{ub}|$ from these branching fractions requires a theoretical
calculation of the form factor, which depends on the $q^2$ range
used. Reducing the $q^2$ range usually improves the error on the form
factor calculation while the experimental error increases because of
the loss of statistics. For $q^2<15\gev^2$ Light Cone Sum
Rules\cite{Ball05} provide the most accurate calculation, whereas
lattice calculation are limited to $q^2>15\gev^2$ due
to the restriction to $\pi$ energies smaller than the inverse lattice
spacing. Using the FNAL04 lattice calculations\cite{FNAL04} for
$q^2>16\gev^2$ one obtains
$$
|V_{ub}|_{\rm excl.} = (\berr{3.75}{0.27}{0.64}{0.42})\times 10^{-3}.
$$

It should be noted that the inclusive and exclusive determinations of
$|V_{ub}|$ are experimentally and theoretically independent. The
previously reported hints of discrepancy\cite{summer2004} between the
two measurements are now reduced in size and the results are
compatible.  Theory errors have been progressively reduced and have
broken the 10\% limit for the inclusive measurement.


\section{$B\to\tau\nu$ decay}
\label{sec:btaunu}

In the SM,  the purely leptonic decay $B^+\to \ell^+
\nu$ (charge conjugate modes are implied) proceeds via the
annihilation of the $\overline b$ and $u$ quark into a virtual $W$
boson. Its amplitude is proportional to the product of $|V_{ub}|$ and
the $B$ meson decay constant $f_B$, with a predicted branching fraction
given by:\cite{Harrison:1998yr}
\begin{eqnarray*}
\lefteqn{\BR(B^+\to \ell^+  \nu)  = {} } &&\\
& & \frac{G_F^2 m_B}{8\pi}m_\ell^2\left(1-\frac{m_\ell^2}{m_B^2}\right)^2f_B^2|V_{ub}|^2\tau_B, 
\end{eqnarray*}
where $G_F$ is the Fermi coupling constants, $m_\ell$ and $m_B$ are
the lepton and $B$ meson masses, and $\tau_B$ is the $B^+$ meson
lifetime. The dependance on the lepton mass arises from helicity
conservation, which suppresses the electron and muon channels.  The
branching ratio in the $\tau$ channel is predicted in the SM to be
roughly $10^{-4}$, but physics beyond the SM, such as supersymmetry or
two-Higgs-doublets models could significantly modify the process.
Observation of $B\to\tau\nu$ would allow a direct determination of
$f_B$, which is currently estimated with a 15\% theoretical
uncertainty\cite{Ryan:2001ej} using lattice QCD calculations.
Besides, the ratio of $\BR(B^+\to
\tau^+ \nu)$ to $\Delta M_{B_d}$, the mass difference between heavy
and light $B_d$ mesons, can be used to determine the ratio of
$|V_{ub}|^2/|V_{td}|^2$, constraining an area in the $\rhobar,\etabar$ plane
with small theoretical uncertainties.\cite{Browder:2000qr} Conversely,
from the global CKM fit one can derive\cite{Charles:2004jd} the
constraint $\BR(B^+\to \tau^+ \nu)=(\cerr{8.1}{1.7}{1.3})\times
10^{-5}$.

Due to the presence of at least two neutrinos in the final state, the
$B^+\to \tau^+ \nu$ decay lacks the kinematical constraints that
are usually exploited in B decay searches to reject both continuum and
\BB backgrounds. The strategy adopted is to exclusively identify the other 
$B$ in the event through a semileptonic or hadronic decay, and assign
all the remaining tracks to the signal $B$. The $\tau$ lepton is then
searched in one or three prongs decays, with a maximum of one
$\pi^0$. After applying kinematical cuts and requiring a large missing
mass in the event, the most powerful variable for separating signal
and background is remaining energy ($E_{ECL}$) non associated with
either $B$. Applying a cut $E_{ECL}<0.3\gev$, Belle\cite{Abe:2005bq} finds no
significant eccess of events over the expected backgrounds, that
ranges between 3 and 12 events depending on the $\tau$ decay mode, 
and sets an upper limit 
$\BR(B^+\to \tau^+ \nu) < 1.8\times 10^{-4} \, @ \, 90 \%{\rm C.L.}$. 
\babar\ finds a slightly higher upper limit.

\begin{figure}
\epsfxsize200pt
\figurebox{}{}{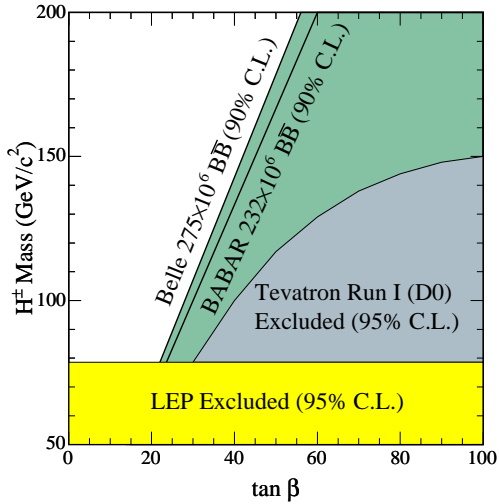}
\caption{Exclusion are in the $[m_H,\tan\beta]$ plane obtained from the upper limit on .$\BR(B^+\to \tau^+ \nu)$.}
\label{fig:tanbeta}
\end{figure}

This result can be interpreted in the context of extensions to the
SM. In the two-Higgs doublet model the decay can occur via
a charged Higgs particle, and the $\BR(B^+\to \tau^+ \nu)$ upper
limited can be translated in a constraint in the $[m_H, \tan\beta]$
plane, as seen in Fig.~\ref{fig:tanbeta} where $m_H$ is the mass of
the Higgs particle and $\tan\beta$ is the ratio of the vacuum
expectation values of the two Higgs doublets.\cite{Hou:1992sy}

\section{$b\to s$ radiative decays}
\label{sec:bsgamma}
Radiative decays involving the $b\to s$ flavour-changing neutral
current transition occur in the SM via one-loop penguin diagrams
containing an up-type quark ($u,c,t$) and a $W$ boson. Example of
these decays are: $B\to X_s\gamma, K^*\gamma, K_S^0\pi^0\gamma,
K\pi\pi\gamma, K^{(*)}\ell^+\ell-, K\nu\nu, \cdots$. 

New physics particles replacing the SM ones in the penguin loop,
e.g. a charged Higgs boson or squarks, can affect both the total rate
of these processes and the decay properties, such as photon
polarization, direct CP violation, and forward-backward asymmetry in
$B\to K^{(*)}\ell^+\ell^-$.

\subsection{$\B\to X_s\gamma$ decays}

Within the SM, the inclusive $\B\to X_s\gamma$ rate is predicted by
next-to-leading order (NLO) calculations\cite{Gambino:2001ew} to be
$\BR(\B\to X_s\gamma)= (3.57\pm0.30)\times 10^{-4}$ for
$E_\gamma>1.6\gev$. The photon energy spectrum provides access to the
distribution function of the $b$ quark inside the $B$
meson,\cite{Neubert:1993um} whose knowledge is crucial for the
extraction of $|V_{ub}|$ from inclusive semileptonic $B\to X_u\ell\nu$
decays, as discussed in Sec.~\ref{sec:vub}.
The heavy quark parameters $m_b$ and $\mu_\pi^2$, which describe the
effective the $b$-quark mass and the kinetic energy inside the $B$
meson, can be determined from the photon energy spectrum, either by
fitting the spectrum directly or by fitting the spectrum
moments.\cite{Neubert:2004sp,Benson:2004sg}

The branching fraction and the photon energy spectrum can be measured
with two methods, originally introduced by CLEO:\cite{Alam:1994aw} in
the fully inclusive method the photon energy spectrum is measured
without reconstructing the $X_s$ system, and backgrounds are
suppressed using event shape variables and high-momentum lepton
tagging of the other $B$; the semi-inclusive method uses a sum of
exclusive final states where possible $X_s$ systems are combined with
the photon, and kinematic constraints are used to suppress
backgrounds.  The semi-inclusive method suffers from uncertainties on
the fragmentation of the $X_s$ system and on the assumptions made as
to the fraction of unmeasured final states. On the other hand the
fully-inclusive method has much larger residual backgrounds that must
be carefully subtracted using off-resonance data. 
\begin{table}
\caption{Summary of partial branching fraction measurements for
the $\B\to X_s\gamma$ process. As explained in the text, 
Belle uses a photon energy cut $E_\gamma>1.8\gev$,
while \babar\ uses $E_\gamma>1.9\gev$.
The errors are statistical, 
systematical, and model dependent.}
\label{tab:bsg}
\begin{center}
\begin{tabular}{|c|c|}
\hline 
Experiment & $\BR(\B\to X_s\gamma) [10^{-4}]$ \\\hline
\raisebox{0pt}[16pt][6pt]{Belle, incl.\cite{Koppenburg:2004fz}} &
\berrsyt{3.55}{0.32}{0.30}{0.31}{0.11}{0.07} \\\hline
\raisebox{0pt}[16pt][6pt]{\babar, incl.\cite{Aubert:2005cb} } &
\err{3.67}{0.29}{0.34\pm0.29} \\\hline
\raisebox{0pt}[16pt][6pt]{\babar, excl.\cite{Aubert:2005cu} } &
\berrsyt{3.27}{0.18}{0.55}{0.40}{0.04}{0.09} \\\hline
\end{tabular}
\end{center}
\end{table}
Table~\ref{tab:bsg} summarizes the $\BR(\B\to X_s\gamma)$
measurements.  Belle uses a photon energy cut $E_\gamma>1.8\gev$,
while \babar\ uses $E_\gamma>1.9\gev$. The results are fully
consistent with the SM expectations. 

\subsection{Photon polarization}
In the SM, the photon from the $b\to s\gamma$ ($\overline b\to
\overline s\gamma$) decays has an almost complete left-handed
(right-handed) polarization. This pattern was generally assumed to be
valid up to a $O(m_s/m_b)$ correction,\cite{Atwood:1997zr} but it has
been recently shown\cite{Grinstein:2004uu} that the corrections can be
significantly larger.  A different polarization pattern would be a
marker of new physics, and can be explored in different ways. In one
method\cite{Atwood:1997zr} photon helicity is probed in mixing-induced
CP asymmetries, exploiting the fact that left-handed and right-handed
photons cannot interfere, thus suppressing time-dependent CP
asymmetries in decays such as $B\to K^{*0}\gamma$.  In another
method\cite{Gronau:2001ng} one uses the kaon resonances decays $B\to
K_{\rm res}\gamma \to K\pi\pi\gamma$ to measure the up-down asymmetry
of the photon direction relative to the $K\pi\pi$ decay
plane. Experimentally, many $B \to K\pi\pi\gamma$ decay channels have
been observed,\cite{Aubert:2005xk,Yang:2004as} with branching
fractions varying in the range $(1.8 - 4.3)\times 10^{-5}$, although
with a statistics still insuffient for the helicity analysis.  Both
decays $B\to K^{*0}\gamma$ and $B\to K_S\pi^0\gamma$ have been
observed and their time-dependent CP asymmetry
measured.\cite{Ushiroda:2005sb,Aubert:2005bu} With the present
statistics all the results are consistent with zero.

\subsection{$B\to K^{(*)}\ell^+\ell^-$ decays}
\begin{figure}
\epsfxsize200pt
\figurebox{}{}{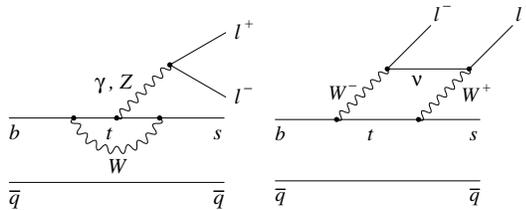}
\caption{Feynmann diagrams decribing the $B\to s\ell^+\ell^-$ decay.}
\label{fig:klldiagram}
\end{figure}

As shown in Fig.~\ref{fig:klldiagram}, $b\to s\ell^+\ell^-$ decays
proceed in the SM both via a radiative penguin diagram with a photon
or a $Z$, and via a $W$-mediated box diagram. The magnitude of the
photon penguin amplitude is known from the $b\to s\gamma$ rate
measurement, while the $Z$ penguin and $W$ box amplitudes provide
new information on FCNC processes. The predicted total branching
fraction is\cite{Ali:2002jg} 
$\BR(b\to s\ell^+\ell^-)=(4.2\pm0.7)\times10^{-6}$, 
in agreement with measurements.\cite{Abe:2004sg,Aubert:2004it}

The $B\to K^{(*)}\ell^+\ell^-$ exclusive decays are predicted to have
branching fractions of $0.4\times10^{-6}$ for $B\to K\ell^+\ell^-$ and
about $1.2\times10^{-6}$ for $B\to K^{*}\ell^+\ell^-$, with a
theoretical uncertainty of about 30\% mainly due the lack of precision
in predicting how often the $s$ quark will result in a single $K^{(*)}$
meson in the final state. Since the electroweak couplings to electron
and muon are identical, the ratio $R_{K}=\BR(B\to
K\mu^+\mu^-)/\BR(B\to Ke^+e^-)$ is expected to be unity, while in
$B\to K^{*}\ell^+\ell^-$ decays a phase space contribution from a pole
in the photon penguin amplitude at $q^2=m_{\ell^+\ell^-}^2\simeq 0$
enhances the lighter lepton pair, with a prediction of
$R_{K^*}=\BR(B\to K^*\mu^+\mu^-)/\BR(B\to
K^*e^+e^-)=0.752$. Neglecting the pole region ($q^2<0.1\gev^2$) for
$B\to K^*e^+e^-$, both ratios $R_K$ and $R_{K^*}$ are predicted to be
very close to unity. However, an enhancement of order 10\% is
expected in the presence of a supersymmetric neutral Higgs boson with
large $\tan \beta$.\cite{Hiller:2003js} New physics at the
electro\-weak scale could also enhance direct CP asymmetries, defined
as $A_{CP}= \frac{\Gamma(\overline B\to K^{(*)}\ell^+\ell^-) -
\Gamma(B\to K^{(*)}\ell^+\ell^-)} {\Gamma(\overline B\to
K^{(*)}\ell^+\ell^-) + \Gamma(B\to K^{(*)}\ell^+\ell^-) }$, to values
of order one,\cite{Kruger:2000zg} while the SM
expectations\cite{Kruger:1999xa} are much less than 1\%. Finally, the
$q^2$-dependance of the lepton forward backward asymmetry is sensitive
to some new physics effects, such as a change of
sign\cite{Hewett:1996ct} of the Wilson coefficient $C_7$ of the
Operator Product Expansion, that would not show up in other channels.
 
\begin{figure}
\epsfxsize200pt
\figurebox{}{}{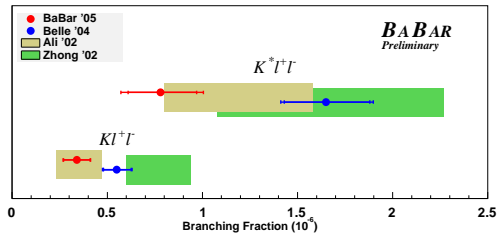}
\caption{Experimental measurements (points) and theoretical predictions for 
$B\to K^{(*)}\ell^+\ell^-$ branching fractions. Red (upper) points are the
\babar\protect\cite{Aubert:2005cf} 
result, while blue (lower) points are the Belle\protect\cite{Abe:2004ir} 
result. The width of the boxes indicates the
estimated precision of the predictions.\protect\cite{Ali:2002jg,Zhong:2002nu}
}
\label{fig:kllsummary}
\end{figure}

Experimentally, the $B\to K^{(*)}\ell^+\ell^-$ decays are identified
through kinematical constraints following a positive $K$
identification. Care must be taken to reject dilepton pairs with a
mass consistent with the $J/\psi$ and the $\psi(2S)$, which are
produced abundantly in $B$ decays. Both processes are well
established,\cite{Aubert:2005cf,Abe:2004ir} and the branching
fractions are compared to theoretical calculations in
Fig.~\ref{fig:kllsummary} CP asymmetries measurements are consistente
with zero with an error of $0.25$.  Belle also reports the first
measurement of the lepton forward-backward asymmetry\cite{Abe:2004ir}
and of the ratio of Wilson coefficients\cite{Abe:2005km}
, although the statistical
power is not yet sufficient to identify new physics effects.

\subsection{Other radiative decays}
Several other exclusive $B$ radiative decay modes have been looked at,
searching for deviation from SM expectations. No signal has been found
yet, but some of the limits (given below at 90\% C.L.) are getting
close to the SM values.  $B\to D^{*0}\gamma$ proceeds via a
$W$-exchange diagram and the branching fraction is expected to be
around $10^{-6}$ in the SM. The measured limit\cite{Aubert:2005aj} is
$\BR(\Bzb\to D^{*0}\gamma)<2.5\times10^{-5}$. For $B\to \phi\gamma$,
which proceeds through a penguin annihilation
diagram\cite{Aubert:2005qc} the SM expectations are around $10^{-12}$,
while the experimental limit is 
$\BR(\Bz \to \phi\gamma)<8.5\times10^{-7}$.  The double radiative decay
$B\to\gamma\gamma$ has a clean experimental signature and is expected
to be around $3\times10^{-8}$ in the SM. The measurements\cite{Abe:2005bs} 
limit its rate at
$\BR(\Bz \to \gamma\gamma)<5.4\times10^{-7}$

\section{Observation of $b\to d$ radiative decays}
\label{sec:bdgamma}


The $b\to d\gamma$ process
is
suppressed with respect to $b\to s\gamma$ by a factor
$|V_{td}/V_{ts}|^2\simeq 0.04$. Due to the large background from
continuum events, only exclusive modes such as $B^-\to\rho^-\gamma$,
$\Bzb\to\rho^0\gamma$, $\Bzb\to\omega\gamma$ (charge conjugate modes
are implied), have been searched so far. Measurement of these
exclusive branching fractions, which are predicted to be in the range
$(0.9 -  2.7)\times10^{-6}$ in the SM,\cite{Ali:2002kw} gives a
precise determination of $|V_{td}/V_{ts}|$ and provides sensitivity to
physics beyond the SM. Belle reports the first
observation\cite{Abe:2005rj} of these decays, reconstructing the
$\rho$ and $\omega$ with final states with at most one
$\pi^0$. Background rejection is obtained through the use of event
shape variables, vertex separtion, and by tagging the other $B$ in the
event. All the variables are used in an unbinned maximum likelyhood
fit where the $B\to(\rho,\omega)\gamma$ and $B\to K^*\gamma$ yields
are simultaneuosly determined.
\begin{figure}
\begin{tabular}{c}
\epsfxsize200pt
\figurebox{}{}{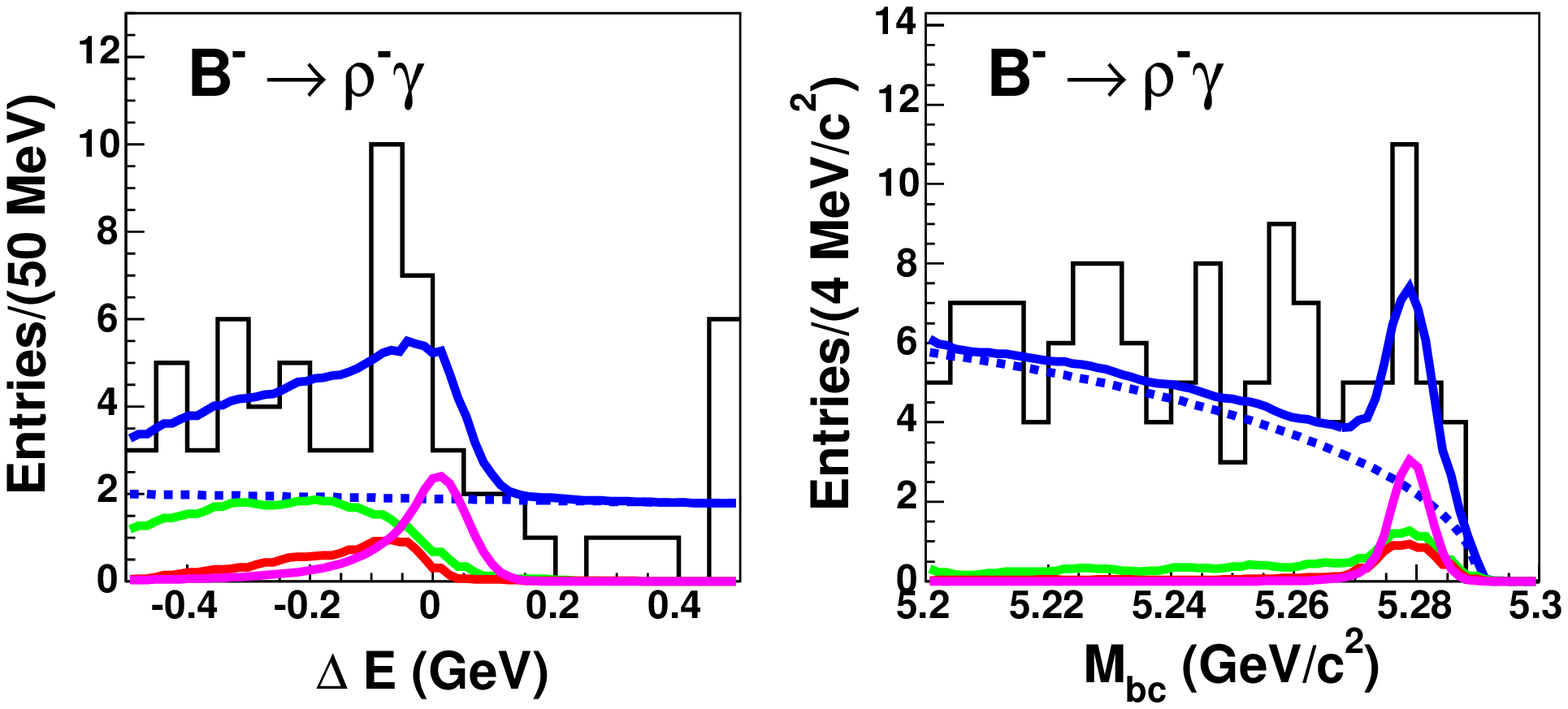}\\
\epsfxsize200pt
\figurebox{}{}{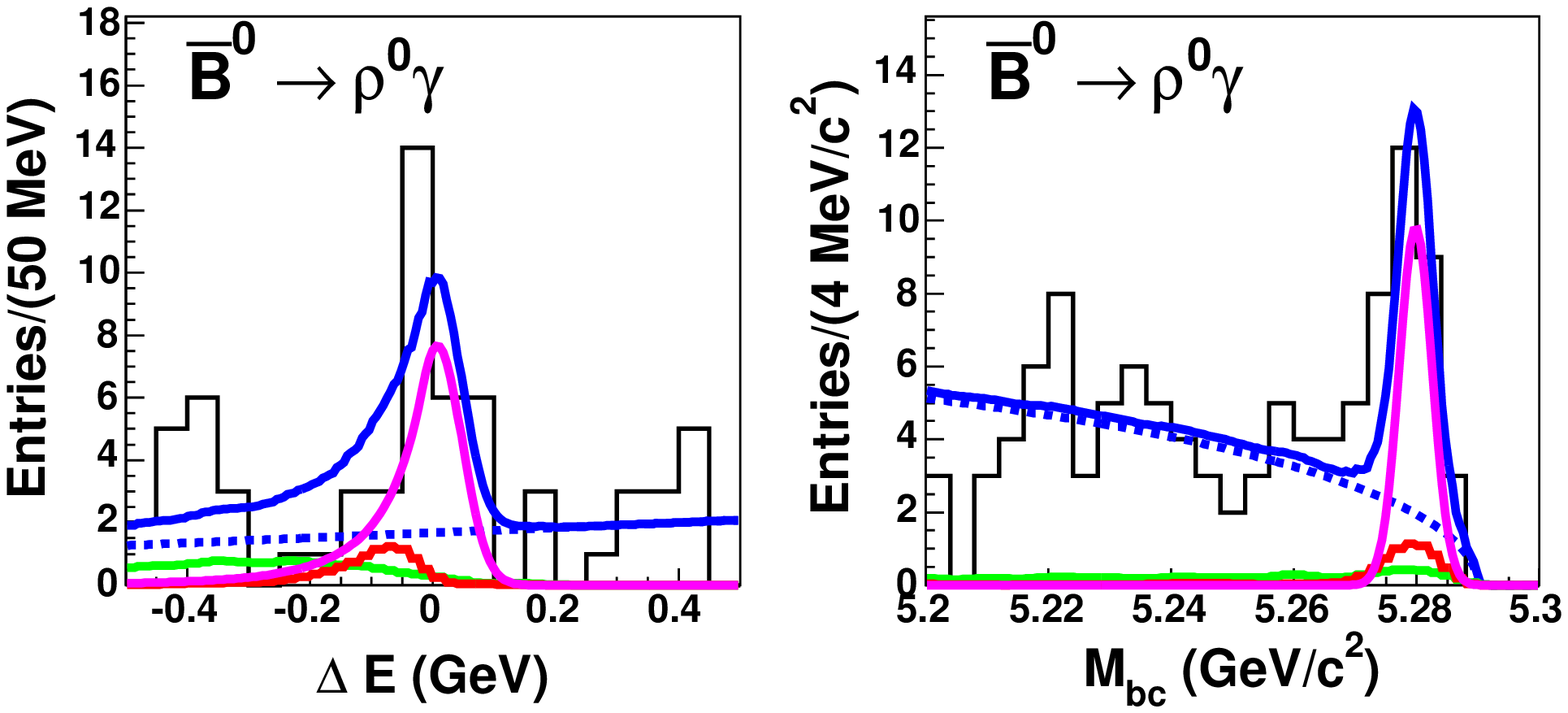}\\
\epsfxsize200pt
\figurebox{}{}{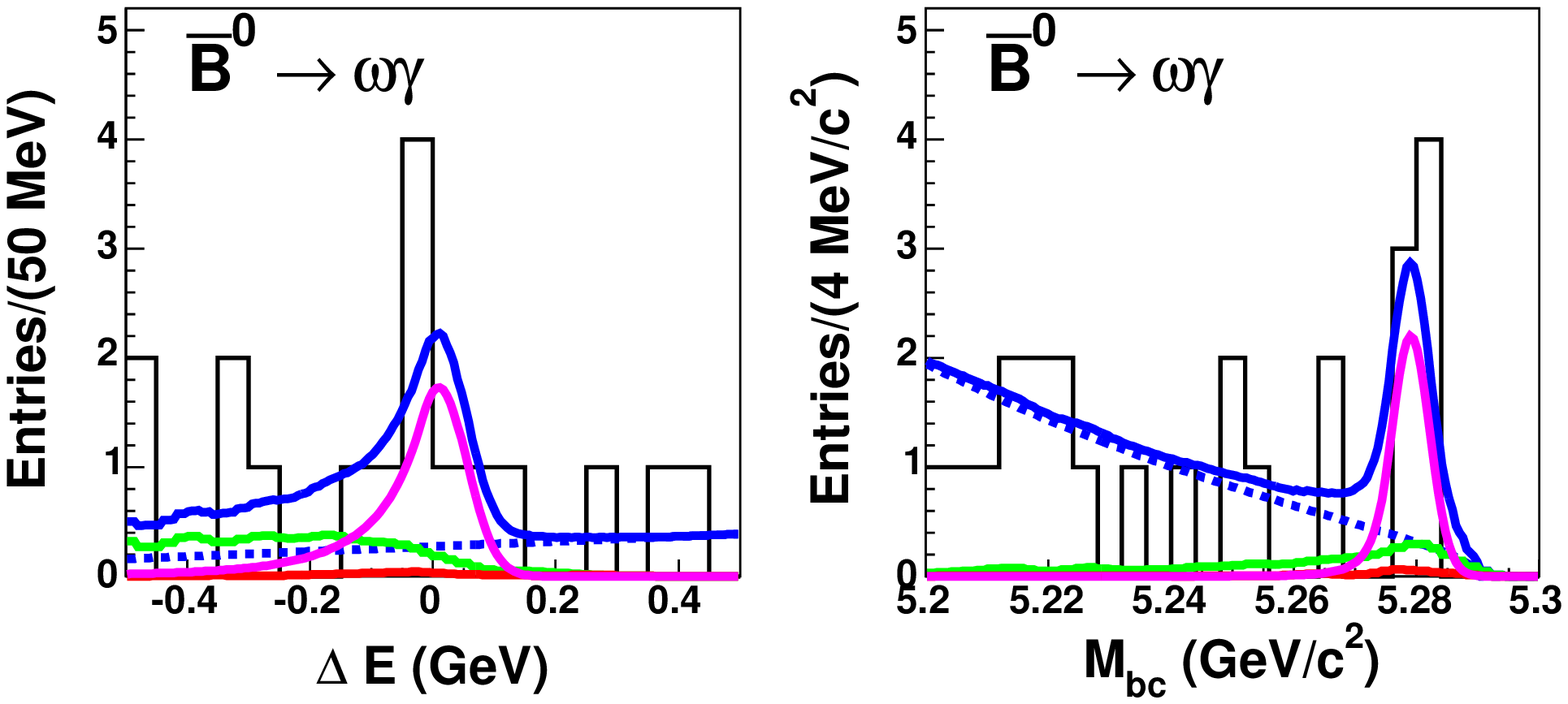}
\end{tabular}
\caption{Projection of the fit results to $M_{bc}$ and $\Delta E$ for the 
    individual $b\to d\gamma$ modes. Lines represent the signal (magenta), continuum (blu-dashed), $B\to K^*\gamma$ (red), other $B$ decay background components (green), and the total fit result (blue-solid). }
\label{fig:dgammapeak}
\end{figure}

Figure~\ref{fig:dgammapeak} shows the projection of the likelihood
fit onto the $M_{bc}$ and $\Delta E$ axes for the individual modes. A
clear peak is always visible. The individual branching ratios are
determined as follows:
\[
\begin{array}{lcl}
\BR(B^-\to \rho^- \gamma) &=& (\aerrsy{0.55}{0.43}{0.37}{0.12}{0.11})\times 10^{-6}, \\[3pt]
\BR(\Bzb\to \rho^0 \gamma) &=& (\aerrsy{1.17}{0.35}{0.31}{0.09}{0.08})\times 10^{-6}, \\[3pt]
\BR(\Bzb\to \omega \gamma) &=& (\aerrsy{0.58}{0.35}{0.27}{0.07}{0.11})\times 10^{-6}, \\[3pt]
\end{array}
\]
where the first error is statistical and the second error is
systematical. The significance figures of the three measurements are
$1.5\sigma, 5.1\sigma$, and $2.6\sigma$, respectively.
A simultaneous fit is also perfomerd using the isospin
relation:
\begin{eqnarray*}
\BR(B\to\rho/\omega\gamma) & \equiv & 
\BR(B^-\to\rho^-\gamma) = 
\\
2 \frac{\tau_{B^+}}{\tau_{B^0}}\BR(\Bzb\to \rho^0\gamma) 
& = & 2 \frac{\tau_{B^+}}{\tau_{B^0}}\BR(\Bzb\to \omega\gamma)
\end{eqnarray*}
where $\tau_{B^+}/\tau_{B^0} = 1.076\pm0.008$ is the ratio of charged
$B$ lifetime to the neutral $B$ lifetime, yielding
\[
\BR(B\to \rho/\omega \gamma) = \aerrsy{1.34}{0.345}{0.31}{0.14}{0.10} \; (5.5\sigma) \\[3pt]
\]
It should be noted that the individual fit results (especially $\BR(B\to
\rho^0 \gamma)$) are in marginal agreement with the isospin relation
above or with the previous limits. More statistics will hopefully
clarify the issue.  The simultaneous determination of $\BR(B\to
K^*\gamma)$ allows the determination of $|V_{td}/V_{ts}|$:\cite{Ali:2004hn}
\[
|V_{td}/V_{ts}|= \aerrsy{0.200}{0.026}{0.025}{0.038}{0.029},
\]
where the errors are respectively from experiment and theory.  This
value is in agreement with global fit to the unitarity
triangle,\cite{Charles:2004jd} but the $b\to d\gamma$ observation
provides an independent constraint on the unitarity triangle which
will become more and more effective as statistics increase.

\section{Summary and conclusions} 
\label{sec:summary}

\begin{figure}
\epsfxsize200pt
\figurebox{}{}{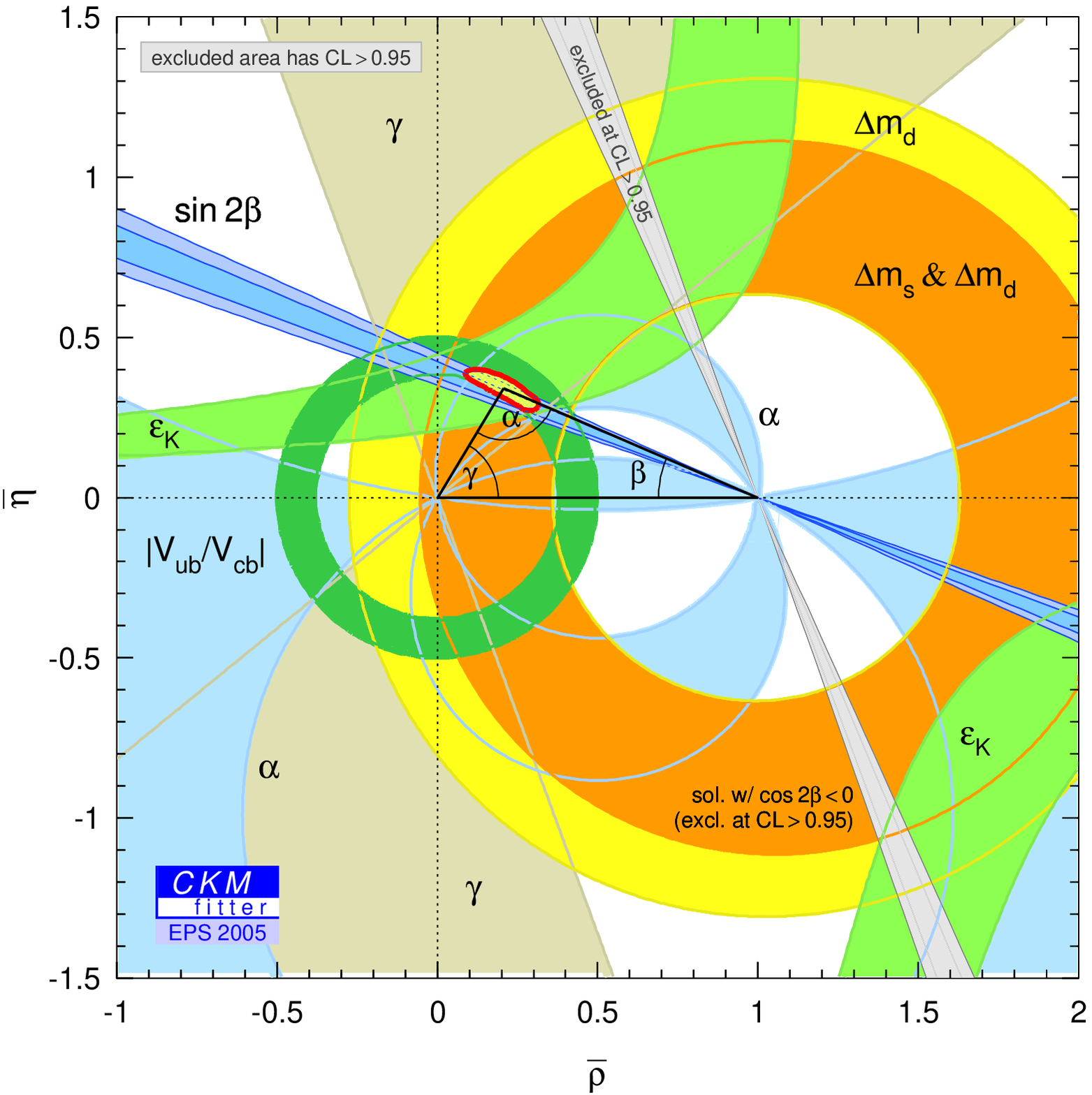}
\caption{Allowed region in the $\rhobar,\etabar$ plane once all the constraints are included.}
\label{fig:rhoeta}
\end{figure}

The accuracy of the analyses performed by the \babar\ and Belle
experiments has been steadily improving, and the precision measurement
of CKM parameters is now a reality. The direct $\alpha(\Phi_2)$
determination $\alpha=(\cerr{99}{12}{9})^\circ$ is for the first time
more precise than its indirect determination from the CKM triangle
fit. $|V_{cb}|$ is known at the 1.5\% level, while $|V_{ub}|_{\rm
incl.} = (\err{4.39}{0.20_{\rm exp}}{0.27_{\rm th}})\times 10^{-3}.$
is determined at the 8\% level, and is the object of intense activity to
further reduce the error.  

Rare decays are very powerful tools for testing the consistency of the
SM and are sensitive to new physics particles in the loop. They also
allow the investigation of the inner structure of the $B$ meson, thus
reducing theory uncertainties in many measurements. $b\to d\gamma$
penguin transitions have been observed at the $5.5\sigma$ level,
$\BR(B\to \rho/\omega \gamma) =
\aerrsy{1.34}{0.345}{0.31}{0.14}{0.10}$, starting to provide new constraints on the unitarity triangle.

Figure~\ref{fig:rhoeta} shows the allowed region in the $\rhobar,\etabar$
plane after all the constraints have been applied. The figure represents the
experimental situation after the summer conferences 2005. As more data will
be necessary to disentangle all the effects and identify the signals
of new physics, each experiment is set to reach a data sample of about
$1~ab^{-1}$ within a few years. Larger samples will require
significant machine and detector upgrades which are being actively
studied by the community.

\section*{Acknowledgments}
I wish to thank my colleagues in Babar and Belle who are responsible
for most of the experimental results presented here. In particular I
am grateful to Vera Luth, Christofer Hearty, Jeff Richman, Marcello
Giorgi, David MacFarlane, Giancarlo Piredda, Riccardo Faccini, Bob
Kowalewski, Jeffrey Berryhill, Iain Stewart, and Yoshi Sakai, for
their comments and help in preparing the talk and this document. A
special thank to Maurizio Pierini and Andreas Hoecker for producing
the UTFit and CKMFitter plots only hours after the results have been
available.  This work has been in part supported by a grant from the
Italian National Institute for Nuclear Physics (INFN).


\end{document}